\documentclass[conference]{IEEEtran}
\IEEEoverridecommandlockouts
\usepackage{cite}
\usepackage{amsmath,amssymb,amsfonts}
\usepackage{algorithmic}
\usepackage{graphicx}
\usepackage{textcomp}
\usepackage{xcolor}
\usepackage{mathrsfs}
\usepackage{amsthm,amsmath,amssymb}
\usepackage{booktabs}
\usepackage{caption}
\usepackage{multirow}
\def\BibTeX{{\rm B\kern-.05em{\sc i\kern-.025em b}\kern-.08em
    T\kern-.1667em\lower.7ex\hbox{E}\kern-.125emX}}
\begin{document}
\captionsetup[figure]{labelfont={rm},labelformat={default},labelsep=period,name={Figure}}
\title{Joint Optimization of Spectrum and Energy Efficiency Considering the C-V2X Security: A Deep Reinforcement Learning Approach\\
}

\author{\IEEEauthorblockN{Zhipeng Liu$^\#$, Yinhui Han$^\#$, Jianwei Fan$^*$, Lin Zhang$^\#$, Yunzhi Lin$^*$}
\IEEEauthorblockA{\textit{$^\#$Beijing University of Posts and Telecommunications, Beijing, China}\\
\textit{$^*$China Railway Electrification Bureau Group Co., Ltd, Beijing, China}\\
\{liuzp0408, \_\_hanyh\_\_, zhanglin\}@bupt.edu.cn, 617096347@qq.com, yunzhilin@hotmail.com}
}

\maketitle

\begin{abstract}
    Cellular vehicle-to-everything (C-V2X) communication, as a part of 5G wireless communication, has been considered one of the most significant techniques for Smart City. Vehicles platooning is an application of Smart City that improves traffic capacity and safety by C-V2X. However, different from vehicles platooning travelling on highways, C-V2X could be more easily eavesdropped and the spectrum resource could be limited when they converge at an intersection. Satisfying the secrecy rate of C-V2X, how to increase the spectrum efficiency (SE) and energy efficiency (EE) in the platooning network is a big challenge. In this paper, to solve this problem, we propose a \underline{S}ecurity-\underline{A}ware Approach to \underline{E}nhancing S\underline{E} and E\underline{E} Based on \underline{D}eep Reinforcement Learning, named SEED. The SEED formulates an objective optimization function considering both SE and EE, and the secrecy rate of C-V2X is treated as a critical constraint of this function. The optimization problem is transformed into the spectrum and transmission power selections of V2V and V2I links using deep Q network (DQN). The heuristic result of SE and EE is obtained by the DQN policy based on rewards. Finally, we simulate the traffic and communication environments using Python. The evaluation results demonstrate that the SEED outperforms the DQN-wopa algorithm and the baseline algorithm by 31.83\% and 68.40\% in efficiency. Source code for the SEED
    is available at https://github.com/BandaidZ/OptimizationofSEandEEBasedonD\\RL.
\end{abstract}


\section{Introduction}
The smart city integrates communication technology and physical devices to optimize the efficiency of city operations and services. Intelligent Transportation System (ITS) is an aspect of Smart City and achieves efficient transportation by V2X communications. As a key technology in the 5G wireless communication \cite{b1}, C-V2X is extended from cellular mobile communication, which includes vehicle-to-vehicle communication (V2V), vehicle-to-infrastructure communication (V2I), vehicle-to-pedestrian communication (V2P) and vehicle-to-network communication (V2N). 

Recently, C-V2X communication has been developed as an important technology in ITS to assist vehicles in sensing each other. And the 3rd Generation Partnership Project (3GPP), a mobile industry body, has introduced some V2X operation scenarios and technical support for them \cite{b2}, such as vehicles platooning. C-V2X communication is used to synchronize their manoeuvres \cite{b3}. Information exchanged in the platoon is conveyed in Cooperative Awareness Message (CAM) periodically, and the update frequency must be at least 10 Hz. With the help of platooning, road safety and traffic efficiency are improved exceedingly.

Platooning is a cooperative driving application where vehicles travelling on the same line keep a small constant inter-vehicle distance to achieve safe transport and to reduce fuel consumption. In the smart city, vehicles platooning not only travel on highways, but also converge at intersections. Different from the highways scenario, it is resource-limited and less secure when they are at intersections because of the high-density vehicles.

Considering the broadcast nature of vehicular platooning communication, it’s easy for malicious users to eavesdrop on other vehicles to achieve all kinds of improper purposes. It is necessary to guarantee the secrecy transmission rates of vehicles, although C-V2X provides excellent convenience for Internet of Vehicles (IoV).

C-V2X has dedicated spectrum and sharing spectrum two modes, operating in 5.9 GHz and 2 GHz, respectively. Spectrum sharing technology is usually used to improve system spectrum efficiency, although it can cause mutual interference between V2V and V2I links. More specifically, in a subframe, the spectrum is divided into several subchannels consisting of a certain number of resource blocks (180 kHz per RB) in the frequency domain. Spectrum sharing means different links use the same subchannels at the same time. In addition, with the development of green communication, energy efficiency becomes more and more attractive to industry and academia in recent years \cite{b4}, since it reveals the relationship between power consumption and the information transmission rate. In a word, these metrics are accurately used to measure the system resource utilization like SE and EE.

\begin{figure*}[htbp]
    \centerline{\includegraphics[scale=0.25]{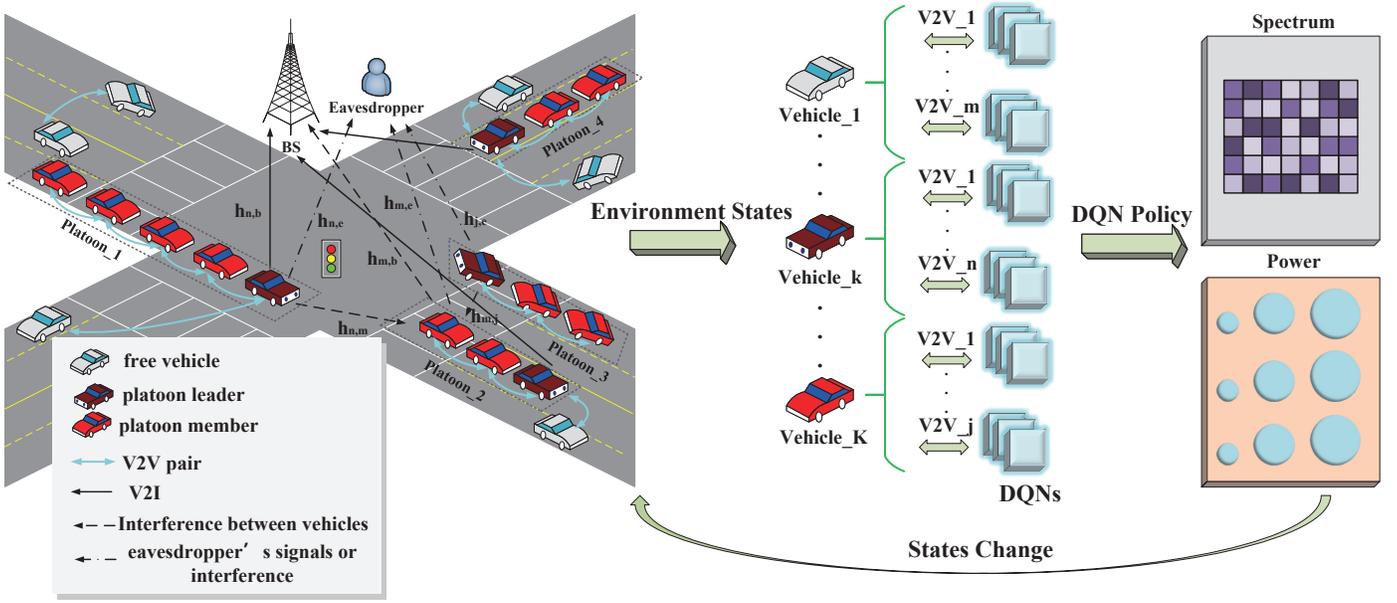}}
    \caption{System model of the SEED.}
    \label{fig}
\end{figure*}

In recent years, there have been some articles about studying the efficiency problem in vehicular communication using C-V2X. \cite{b5} uses the upper bound of the outage probability to make V2V links reuse maximization problem. It falls under the framework of the Perron-Frobenius Theorem to improve spectrum utilization. \cite{b6} investigates the secure radio resource sharing problem in device-to-device (D2D) enabled multi-platooning vehicular communication, and they propose a security-aware joint channel and power allocation (SA-JCPA) scheme using a two-step approach. \cite{b7} considers stringent latency constraints on V2V links and proposes a decentralized deep reinforcement learning method to enhance spectrum efficiency in both unicast and broadcast scenarios. In terms of security and privacy, \cite{b8} proposes a fully uncoordinated approach to change pseudonyms in distributed networks, where each node uses a pseudonym until its expiration and then changes after a random delay, and the analysis confirmes that the k-anonymity can be achieved at a negligible throughput loss in the case of large networks.

However, most articles in C-V2X focus on either SE or EE without thinking about the secrecy transmission rates. To be more comprehensive, we optimize both of them jointly using the SEED, a security-aware approach based on deep reinforcement learning in the scenario where vehicles platooning exist. The main contributions in this paper are listed as follows.

\begin{itemize}
    \item To reflect the efficiency of vehicular network systematically, spectrum and energy efficiency on two different types of V2V and V2I links are considered. The efficiency of V2V and V2I links is weighted in the light of their using frequency. To solve the explosion of state and action in traditional reinforcement learning, DQN is used to optimize the problem intelligently.
    \item To the best of our knowledge, this is the first attempt to solve the spectrum and energy efficiency of vehicular platooning network problem that takes into account the secrecy rates of V2V links based on physical layer security.
    \item According to the characteristics of the transmission power of V2V and V2I links in vehicles platooning, platooning leaders and members can be allocated dedicated transmission power based on DQN policies.
\end{itemize}

The rest of this paper is organized as follows. Section II describes the system model and presents the joint optimization analysis of both SE and EE considering the secrecy rates of V2V links, then an approach named SEED solves it heuristically. Simulation results are discussed in Section III. Eventually, our conclusions are summarized in Section IV.

\section{System Model}
In this section, the system model is presented. As depicted in figure 1, there is a dense and resource-limited vehicles platooning scenario where an eavesdropper exists. In this scenario, multi-platooning and free vehicles gather in an intersection. In a platoon, platooning leader tries to transmit CAMs to all the platooning members, and each platooning member transmits its own CAM to the neighbors. Platooning leaders and free vehicles get information about traffic and entertainment by V2I communications. The resource allocation management mechanism applied in this article is C-V2X \emph{mode-4}. In C-V2X \emph{mode-4}, vehicles select radio spectrum resources in a distributed manner. Each V2V link, as an agent, has a DQN model to determine which subchannel and what power level to choose. Then they jointly decide the spectrum and power level selections in the platooning network by a shared reward function. Details will be provided in Part \emph{C}.

Assuming that the orthogonal frequency division multiplexing
(OFDM) technique is employed to support V2I communications and multi-platooning V2V communications. Assuming at one time scheduling unit (a subframe), the vehicular platooning network includes \emph{M} V2V links, denoted as $\mathcal{M}$ = \{1,2,...,\emph{m},...,\emph{M}\}, \emph{N} V2I links, denoted as $\mathcal{N}$ = \{1,2,...,\emph{n},...,\emph{N}\} and \emph{N} subchannels. 
Considering the lower usage of V2I links compared to V2V links, to solve problems conveniently, the number of V2I links is constant \emph{N} in a subframe. V2V UEs share CAMs with each other using V2V links, and V2I UEs use V2I links to communicate with road infrastructure like base station (BS). We assume that V2V links reuse the uplink spectrum allocated orthogonally to V2I links since the interference at the BS is more controllable.

\subsection{Communication Model}\label{AA}
The subchannels multiplexing allocation matrix is
\begin{subequations}
    \begin{gather}
        {\bf{A}} = \left[ {\begin{array}{*{20}{c}}
            {{a_{11}}}&{{a_{12}}}& \cdots &{{a_{1N}}}\\
            {{a_{21}}}&{{a_{22}}}& \cdots &{{a_{2N}}}\\
             \vdots & \vdots & \vdots & \vdots \\
            {{a_{M1}}}&{{a_{M2}}}& \cdots &{{a_{MN}}}
            \end{array}} \right] \\
            {a_{mn}} = \left\{ \begin{array}{l}
                1,m{\rm{th \,V2V \,link \,reuse \,}}n{\rm{th \,V2I \,link}}\\
                0,otherwise
                \end{array} \right.
    \end{gather}

\end{subequations}

The transmission rate of \emph{n}th V2I link can be expressed as

\begin{equation}
    R_n^{} = {\log _2}(1 + \frac{{{P_{V2I}}{h_{n,b}}}}{{\sum\limits_{m = 1}^M {{a_{mn}}{P_{V2V}}{h_{m,b}} + {\sigma ^2}} }})
\end{equation}
where ${P_{V2I}}$ and ${P_{V2V}}$ mean the transmission power of \emph{n}th V2I link and \emph{m}th V2V link, respectively. ${a_{mn}}$ means the subchannels multiplexing allocation element. ${h_{n,b}}$ and ${h_{m,b}}$ denote the power gains of the channel from \emph{n}th V2I link to the BS and from \emph{m}th V2V link using the same subchannel with \emph{n}th V2I link to the BS, respectively. ${\sigma ^2}$ is the noise power. And the transmission rate of \emph{m}th V2V link can be expressed as

\begin{subequations}
    \begin{gather}
        R_m^{} = {\log _2}(1 + \frac{{{P_{V2V}}{h_{m}}}}{{{I_m} + {\sigma ^2}}}) \\
        {I_m} = \sum\limits_{n = 1}^N {{a_{mn}}{P_{V2I}}{h_{n,m}} + \sum\limits_{n = 1}^N {\sum\limits_{j \ne m}^M {{a_{mn}}{a_{jn}}{P_{V2V}}{h_{m,j}}} } }
    \end{gather}
\end{subequations}
where ${h_{m}}$ and ${h_{m,j}}$ are the power gain of the channel corresponding to \emph{m}th V2V link and the interference power gain of \emph{j}th V2V link. ${I_m}$ means the interference to \emph{m}th V2V link. Other elements are the same as mentioned above.

For the eavesdropper, the transmission rate of eavesdropping on a V2V link is

\begin{subequations}
    \begin{gather}
        R_{m,e}^{} = {\log _2}(1 + \frac{{{P_{V2V}}{h_{m,e}}}}{{{I_{eve}}{\rm{ + }}{\sigma ^2}}}) \\
        {I_{eve}} = \sum\limits_{n = 1}^N {{a_{mn}}{P_{V2I}}{h_{n,e}} + \sum\limits_{n = 1}^N {\sum\limits_{j \ne m}^M {{a_{mn}}{a_{jn}}{P_{V2V}}{h_{j,e}}} } }
    \end{gather}
\end{subequations}    
where ${h_{m,e}}$, ${h_{n,e}}$ and ${h_{j,e}}$ are the power gain of channel eavesdropping on other V2V links, the interference power gain of \emph{n}th V2I link and \emph{j}th V2V link, respectively, and $I_{eve}$ is the interference to the eavesdropper. The concept of secrecy rate is the non-negative difference between transmission rate and eavesdropping rate. The secrecy rate of V2V links are

\begin{equation}
    {R_{m,\sec urity}} = {[{R_m} - {R_{m,e}}]^ + }
\end{equation}
where y = ${[x]^ + }$ means 

\begin{equation}
    {\rm{y}} = \left\{ \begin{array}{l}
        x,x \ge 0\\
        0,x < 0
        \end{array} \right.
\end{equation}

\subsection{Problem Formulation}
In a dense and resource-limited traffic environment like intersections, there is a marked drop in the SE and EE of V2V and V2I links. And there may be users called eavesdroppers, who illegally monitor vehicular information and keep records of it to get their privacy. This article optimizes the vehicular platooning network resource utilization including both SE and EE under the premise of ensuring the V2V secrecy rates.

Multi-objective optimization problems with different dimensions need to figure out the Pareto optimality. To optimize the problem efficiently and comprehensively, a compositive efficiency is defined as the ratio of SE and the total power consumption, which integrates the meaning of SE and EE \cite{b9}. The compositive efficiency including SE and EE of V2V links is

\begin{equation}
    {\zeta _{V2V}}{\rm{ = }}\sum\limits_{m = 1}^M {{B_n}{R_m}} /(\sum\limits_{n = 1}^N {{B_n}} *(\sum\limits_{m = 1}^M {{P_{V2V}} + M{P_C}} ))
\end{equation}
where ${P_C}$ is the circuit power, ${B_n}$ means the bandwidth, and the compositive efficiency of V2I links is

\begin{equation}
{\zeta _{V2I}} = \sum\limits_{n = 1}^N {{B_n}{R_n}} /(\sum\limits_{n = 1}^N {{B_n}} *(\sum\limits_{n = 1}^N {{P_{V2I}}}  + N{P_C}))
\end{equation}

Overall, the objective optimization function can be summarized as

\begin{equation}
max {\rm{ }}\,{\lambda _\alpha}{\zeta _{V2V}} + {\lambda _\beta }{\zeta _{V2I}}
\end{equation}
subject to:
\begin{subequations}
\begin{align}
\sum\limits_{n = 1}^N {{a_{mn}} \le 1} ,\forall m \\
{R_{m,\sec urity}} \ge {R_T},\forall m \\
0 \le {P_{V2I}} \le {P_{\max 1}} \\
0 \le {P_{V2V}} \le {P_{\max 2}} \\
{a_{mn}} = \{ 0,1\}
\end{align}
\end{subequations}

(9) is the objective optimization function to maximize the SE and EE of V2V and V2I links. (10a) demonstrates \emph{m}th V2V link can most reuse one V2I link. (10b) means the constraint of the V2V secrecy rate. (10c) and (10d) limit the V2I and V2V transmission power. (10e) is the subchannels multiplexing allocation element limited on 0 or 1.

\subsection{Joint Optimization Analysis of SE and EE Considering the secrecy rates of V2V}
In the Part \emph{B} of Section III, the optimization function (9) is a combinatorial optimization problem with discrete integers. Obviously, the optimization problem is a NP-hard problem, so it is difficult to obtain the optimal solution. Based on these assumptions and models, the SEED is proposed to optimize the resource utilization of V2V and V2I links, which is a security-aware approach to enhancing the spectrum efficiency and energy efficiency based on deep reinforcement learning.

Deep reinforcement learning (DRL) is a combination of deep learning and reinforcement learning. DRL sees learning as a heuristic evaluation process. The main idea of DRL is that an agent may learn by interacting with its environment using the experience gathered. It should be able to optimize some objectives given in the form of cumulative rewards \cite{b10}. Compared to traditional Q learning, deep Q network is an improved algorithm which solves the problem of state and action space explosion. Thus it is used to solve our optimization problem heuristically \cite{b11}. The parameters of DQN are set below in this article’s scenario.

\begin{figure}[htbp]
    \centerline{\includegraphics[scale=0.15]{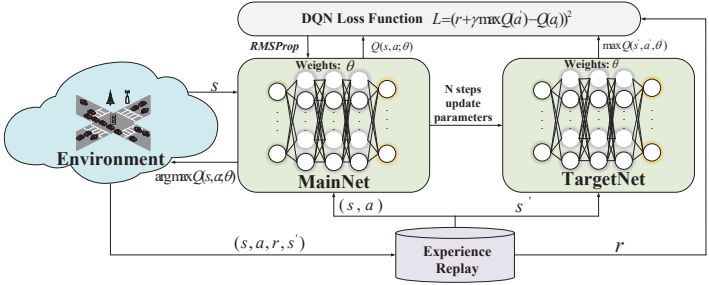}}
    \caption{The structure of DQN in the SEED.}
    \label{fig}
\end{figure}

\begin{itemize} 
    \item\textbf{Agent:} \emph{v} $\in \{ {v_1},{v_2},...,{v_m},...\}$, where \emph{v} is a V2V link existing in the network.

\item \textbf{State Space:} \emph{s} $\in \{ H_{m}^{},H_{n,b}^{},H_{m,e}^{},H_{n,e}^{},I_i^{},N_i^{}\}$, where \emph{s} is the state observed from environment by the agent \emph{v}. ${H_{m}}$ are the subchannel power gains of V2V links. ${H_{n,b}}$ are the subchannel power gains from V2I UEs to BS. ${H_{m,e}}$ and ${H_{n,e}}$ mean are power gains to eavesdroppers. ${I_i}$ are the interferences to the agents. ${N_i}$ are the subchannels selected by local observations.

\item\textbf{Action Space:} \emph{a} $\in \{ \{ {N_1},{P_{V2{V^{(1)}}}}\},\{ {N_2},{P_{V2{V^{(2)}}}}\},...,\\\{ {N_K},{P_{V2{V^{(K)}}}}\} \}$, where ${N_k}$ is the subchannel that the agent selects, ${P_{V2{V^{(K)}}}}$ is the V2V transmission power level.
Vehicles platooning's communication is different from that of free vehicles, because platoon members mainly transmit and receive intra-platoon messages (e.g. intra-platoon CAM) to maintain the internal safety of the platoon. Therefore, it isn't necessary for them to use the transmission power, which closes to the maximum power in the standard protocol of 3GPP.
Thus, the V2V transmission power level of platoon members is lower than that of platoon leaders and free vehicles.

\item\textbf{Reward Function:}
\begin{equation}
    \emph{r} = \left\{ \begin{array}{l}
        {\lambda _\alpha}{\zeta _{V2V}} + {\lambda _\beta }{\zeta _{V2I}},{\rm{ }}if{\rm{ }}{R_{m,\sec urity}} \ge {R_T}\\
        {\rm{ - }}1,{\rm{ }}otherwise
        \end{array} \right.
\end{equation}
\end{itemize}
where ${\lambda _\alpha }$ and ${\lambda _\beta }$ are weighted coefficients representing its importance in the vehicular network, and ${\lambda _\alpha }$ + ${\lambda _\beta }$ = 1.
A negative reward is introduced in order to make DQN select better policies.
Obviously, the threshold of V2V secrecy rate is the boundary of positive and negative rewards, and guarantees that the agent can get more positive rewards if it have a certain number of secrecy rates.
This reward function is shared by each agent.
During the interaction between each agent and the environment, the corresponding actions are continuously updated to obtain the maximum cumulative reward.

The structure of DQN in the SEED is illustrated in figure 2. The framework of DQN consists of agents and the vehicular environment. In a subframe, an agent \emph{v} observes a state \emph{$s_{t}$} from the environment, and the MainNet can get the state-action function $Q(s,a;\theta )$, which is used to evaluate the selection of the action. According to the action the agent selects, it can get a new state \emph{${s_{t+1}}$}. The function $Q({s}',{a}';{\theta }')$ means the output of the TargetNet, and the target Q is $r + \gamma \max Q({s}',{a}',{\theta}')$. So the loss function in the deep neural network is

\begin{equation}
    L = {(r + \gamma \max Q({s}',{a}',{\theta}') - Q(s,a,\theta))^2}
\end{equation}
where \emph{r} is the value of reward when taking action \emph{a}. $\gamma$ is the discount factor, the closer it is to 1, the more it will value future rewards, and the closer it is to 0, the less sensitive it is to future rewards. The asynchronization between parameters ${\theta}'$ and $\theta$ sloves the problem of instability when using nonlinear network to evaluate value function.

\section{Simulations}

In this section, we present simulation setup and simulation results to show the performance of the proposed SEED in the spectrum and energy efficiency under the premise of maintaining a certain number of the V2V secrecy rates.

\subsection{Simulation Setup}
Before the simulation analysis, the basis of parameter selection should be introduced.
The frequency operates in 2 GHz to realize spectrum sharing between V2V and V2I links detailed in 3GPP TR 36.785. One V2V link multiplexes at most one V2I link. We follow the urban scenario in 3GPP TR 37.885. It is 1299 \emph{m} long and 750 \emph{m} wide. The vehicles are placed in the scenario as platooning, and each platooning has five vehicles including one platooning leader and four platooning members. Vehicle speed is set at 60 \emph{km/h}. 

Experimental environment relies on Python. The DQN model is built in TensorFlow 1.9.0 and the GPU model used is GeForce GTX 1080 Ti. The number of hidden layers is 3. Optimizer we select is RMSProp, learning rate is set as 0.01. Discount factor is 0.5. The size of mini batch is 2000. More details are marked in Table I.

\begin{table}[htbp]
\caption{simulation parameters}
\begin{center}
\scalebox{0.68}{
\begin{tabular}{c c|c c}
\toprule[1.5pt]
\textbf{Communication Parameter}&\textbf{Value}&\textbf{Traffic and DQN Parameter}&\textbf{Value} \\
\midrule[1pt]
Carrier Frequency&2 GHz&Number of Vehicles& [20,40,60,80,100]  \\
Total Bandwidth&10 MHz&Size of Platoons&5 \\
Number of Subchannels&20&Platoon Speed&60 km/h \\
V2I Power&23 dBm&Lane Width&3.5 m \\
V2V Power Level&[23,15,10,5] dBm&Number of Lanes&4 in each direction \\
Circuit Power&16 dBm&Number of Direction&4 \\
BS Antenna Gain&8 dBi&Number of Hidden Layers&3 \\
Eavesdropper Antenna Gain&6 dBi&Optimizer&RMSProp \\
Vehicle Antenna Gain&3 dBi&Activation Function&Relu \\
Noise Power&-114 dBm&Batch Size&2000 \\
Decorrelation Distance&10 m&Learning Rate&0.01 \\
Path Loss&LOS NLOS&Discount&0.5 \\
Shadow Fading Distribution&Lognormal&Update Steps&100 \\
\bottomrule[1.5pt]
\end{tabular}}
\label{tab1}
\end{center}
\end{table}

\subsection{Simulation Results of the SEED}

In order to evaluate the performance of the proposed SEED, we compare it with the DQN algorithm without power allocation (DQN-wopa) \cite{b12} and the baseline algorithm that uses random allocation.

\begin{figure}[htbp]
    \centerline{\includegraphics[scale=0.5]{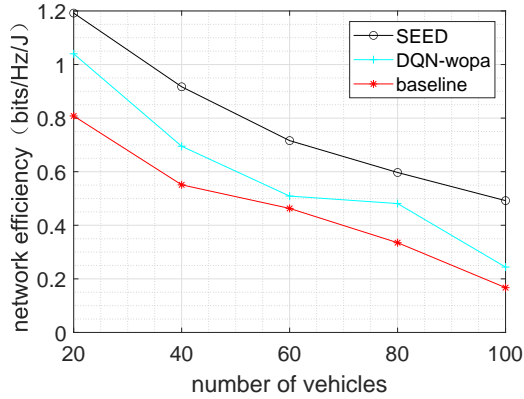}}
    \caption{Network efficiency including SE and EE versus the number of vehicles.}
    \label{fig}
\end{figure}

\begin{figure}[htbp]
    \centerline{\includegraphics[scale=0.5]{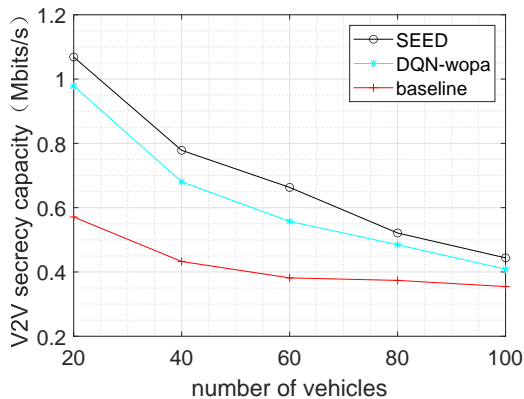}}
    \caption{Mean V2V secrecy capacity versus the number of vehicles.}
    \label{fig}
\end{figure}

As shown in figure 3, with the increase of the vehicles, the network efficiency including SE and EE in both V2V and V2I links is decreasing. The reason is that the power consumption and the interference of the network will be ascending when the number of vehicles increases. From this figure, we can see that the SEED proposed alleviates this phenomenon to a greater extent. The numerical value of network efficiency is 1.192 when the number of vehicles is 20, and the numerical values of DQN-wopa algorithm and the baseline algorithm are 1.040 and 0.626, respectively.
When the number of vehicles is 100, the environment of the system becomes severe, the network efficiency of V2V and V2I links is 0.492 using the SEED proposed, which is better than 0.383 and 0.167 using the other two schemes.
The remaining values in the SEED are 0.917, 0.716 and 0.597.
We can see that the down curve of using SEED is relatively flat, indicating that for the high-density vehicles environment, it can better slow down the decrease in efficiency.

Figure 4 demonstrates the mean V2V secrecy capacity versus the number of vehicles. The SEED and the DQN-wopa algorithm both have higher performance on protecting vehicles' privacy than the baseline algorithm, because the V2V secrecy rate is considered as the constraint of the optimization problem. Because the SEED can choose different power for vehicles adaptively, it will interfere as much as possible with the eavesdropper. While the performance of random algorithm is bad because it does not care about the security of C-V2X.
The values of mean V2V secrecy rates are 1.068, 0.778, 0.663, 0.521 and 0.443 in the SEED.

\begin{figure}[htbp]
    \centerline{\includegraphics[scale=0.5]{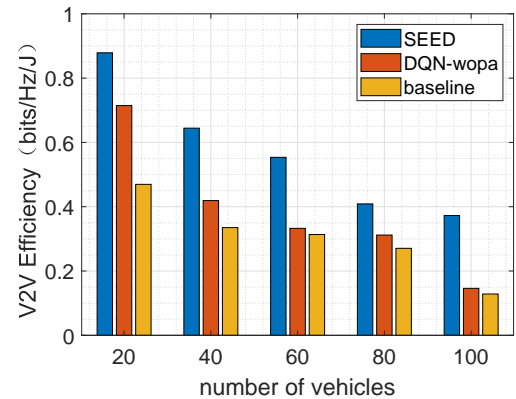}}
    \caption{V2V efficiency versus the number of vehicles.}
    \label{fig}
\end{figure}

\begin{figure}[htbp]
    \centerline{\includegraphics[scale=0.5]{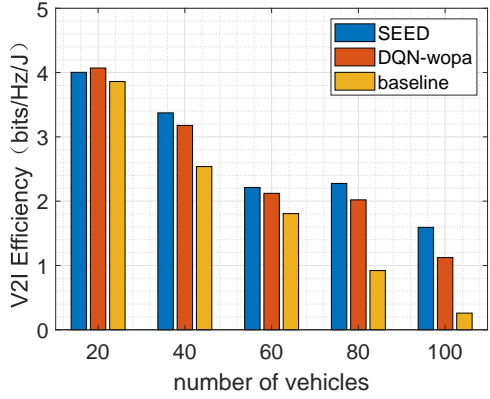}}
    \caption{V2I efficiency versus the number of vehicles.}
    \label{fig}
\end{figure}

Figure 5 and figure 6 show that the network efficiency of V2V links and the network efficiency of V2I links.
As shown in formula (9), the optimization problem in this paper is a combinatorial one.
${\lambda _\alpha}$ and ${\lambda _\beta}$ are set as 0.9 and 0.1, respectively. These weighted coefficients mean that the network efficiency of V2V links is more important than that of V2I links, the reason is that the usage frequency of V2V is more frequent than that of V2I.

From figure 5 we can see that the network efficiency of V2V links in the SEED far superior to the other two schemes. In the SEED, when the number of vehicles is from 20 to 100, the numerical values are 0.879, 0.644, 0.554, 0.409 and 0.372, respectively. The reason is that the network efficiency of V2V links accounts for a large proportion in the optimization function (9).

As shown in figure 6, the network efficiency of V2I links in the approach proposed is almost as good as DQN-wopa algorithm. When the number of vehicles is from 20 to 100, the numerical values are 4.004, 3.373, 2.213, 2.276 and 1.593 in our approach, respectively. The performance of the SEED on the V2I efficiency is not the most excellent in these three schemes every time. The reason is that the weighted coefficient of V2I is small, to maximize the efficiency of the whole network including V2V and V2I links, the SEED makes its choice.

\begin{figure}[htbp]
    \centerline{\includegraphics[scale=0.5]{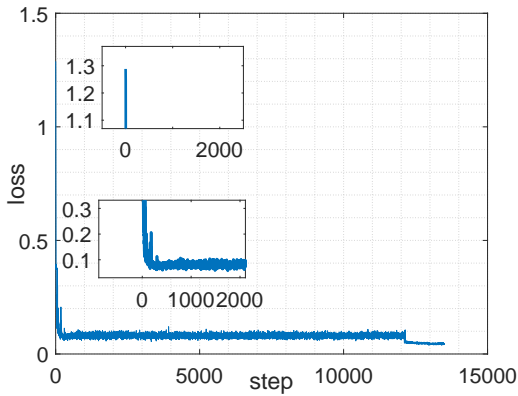}}
    \caption{Loss function.}
    \label{fig}
\end{figure}

The loss function of our DQN model in 20 vehicles is shown in figure 7. It can be seen that when training about 150 steps, the loss value drops from 1.288 to about 0.107. After long time steps, the loss value falls to about 0.047.
It seems that the model has already converged.

\section{Conclusion}

In this paper, we studied how to improve the spectrum and energy efficiency of V2V and V2I links in the dense and resource-limited scenario considering the C-V2X security. To accomplish this aim, a security-aware approach to enhancing the spectrum efficiency and energy efficiency based on deep reinforcement learning for both V2V and V2I links was proposed, named SEED. We not only optimized the spectrum efficiency and energy efficiency but also ensured the V2V secrecy rates. For this purpose, DQN maximized the network efficiency including SE and EE heuristically, and the V2V secrecy rate was considered as a critical constraint. According to the characteristics of the transmission power of V2V and V2I links in vehicles platooning, platooning leaders and members can be allocated dedicated transmission power by DQN policies. The simulation results showed that the SEED outperformed DQN-wopa algorithm and random algorithm in Smart City. In the future research, we will consider the closed-form expression for this problem.

\vspace{12pt}


\begin{thebibliography}{00}
\bibitem{b1} P. Wang, B. Di, H. Zhang, K. Bian and L. Song, "Platoon Cooperation in Cellular V2X Networks for 5G and Beyond," in IEEE Transactions on Wireless Communications, vol. 18, no. 8, pp. 3919-3932, Aug. 2019.
\bibitem{b2} 3GPP. Study on LTE-based V2X Services: TR 36.885 v.14.0.0[S]. 2016.
\bibitem{b3} N. Giovanni, V. Antonio, C. Claudia, M. Antonella and S. Giovanni, "Cellular-V2X Communications for Platooning: Design and Evaluation," Sensors (Basel, Switzerland), vol. 18, no. 5, May 2018.
\bibitem{b4} Y. Yang, Y. Zhang, K. Shi and J. Li, "Optimal power control for energy efficiency of device-to-device communication underlaying cellular networks," 2016 IEEE 14th International Conference on Industrial Informatics (INDIN), Poitiers, 2016, pp. 1028-1031.
\bibitem{b5} Q. Wen, B. Hu and L. Zheng, "Outage-Constrained Device-to-Device Links Reuse Maximization and Its Application in Platooning," in IEEE Wireless Communications Letters, vol. 8, no. 6, pp. 1635-1638, Dec. 2019.
\bibitem{b6} X. Peng, H. Zhou, B. Qian, K. Yu, N. Cheng and X. Shen, "Security-Aware Resource Sharing for D2D Enabled Multiplatooning Vehicular Communications," 2019 IEEE 90th Vehicular Technology Conference (VTC2019-Fall), Honolulu, HI, USA, 2019, pp. 1-6.
\bibitem{b7} H. Ye, G. Y. Li and B. F. Juang, "Deep Reinforcement Learning Based Resource Allocation for V2V Communications," in IEEE Transactions on Vehicular Technology, vol. 68, no. 4, pp. 3163-3173, April 2019.
\bibitem{b8} Z. Liu, L. Zhang, W. Ni and I. Collings, "Uncoordinated Pseudonym Changes for Privacy Preserving in Distributed Networks," in IEEE Transactions on Mobile Computing. doi: 10.1109/TMC.2019.2911279
\bibitem{b9} X. Chen, X. Wu, S. Han and Z. Xie, "Joint Optimization of EE and SE Considering Interference Threshold in Ultra-Dense Networks," 2019 15th International Wireless Communications and Mobile Computing Conference (IWCMC), Tangier, Morocco, 2019, pp. 1305-1310.
\bibitem{b10} G. Cao, Z. Lu, X. Wen, T. Lei and Z. Hu, "AIF: An Artificial Intelligence Framework for Smart Wireless Network Management," in IEEE Communications Letters, vol. 22, no. 2, pp. 400-403, Feb. 2018.
\bibitem{b11} W. Tong, A. Hussain, W. X. Bo and S. Maharjan, "Artificial Intelligence for Vehicle-to-Everything: A Survey," in IEEE Access, vol. 7, pp. 10823-10843, 2019.
doi: 10.1109/ACCESS.2019.2891073
\bibitem{b12} Z. Liu, X. Chen, Y. Chen and Z. Li, "Deep Reinforcement Learning Based Dynamic Resource Allocation in 5G Ultra-Dense Networks," 2019 IEEE International Conference on Smart Internet of Things (SmartIoT), Tianjin, China, 2019, pp. 168-174.
\end{thebibliography}
\end{document}